\documentclass[aps,pre,twocolumn,superscriptaddress]{revtex4}
\usepackage[latin1]{inputenc}
\usepackage{textcomp}
\usepackage{amsmath}
\usepackage{amssymb}
\usepackage{graphicx}
\usepackage{wasysym}
\usepackage{physics}
\usepackage{stackrel}
\usepackage{multirow}
\usepackage{braket,mleftright}
\usepackage{empheq}
\usepackage{subfigure}
\usepackage{soul}
\usepackage{blkarray}
\usepackage{bm}
\usepackage{bbold}
\usepackage{color}
\usepackage[unicode=true,
 bookmarks=false,
 breaklinks=false,pdfborder={0 0 1},backref=false,colorlinks=false]
 {hyperref} 
\usepackage{breakurl}
\usepackage{dsfont}
\makeatletter

\usepackage{amsfonts}
\usepackage{amsthm}
\usepackage{graphics}
\usepackage{times}
\usepackage{bm}
\usepackage{color}
\usepackage{latexsym}
\usepackage{subfigure}
\usepackage{braket}
\usepackage{float}
\usepackage{palatino}
\usepackage{bbm}
\usepackage{color}

\newcommand{\beq}{\begin{equation}}
\newcommand{\eeq}{\end{equation}}
\newcommand{\bse}{\begin{subequations}}
\newcommand{\ese}{\end{subequations}}
\newcommand{\bea}{\begin{eqnarray}}
\newcommand{\eea}{\end{eqnarray}}

\makeatother

\begin{document}

\title{Quantum work for sudden quenches in Gaussian random Hamiltonians}

\author{Eric G. Arrais}
\email{eric.arrais@gmail.com}
\affiliation{Instituto de F\'{i}sica, Universidade Federal do Rio de Janeiro, 21941-972, Rio de Janeiro, Brazil}

\author{Diego A. Wisniacki}
\affiliation{Departamento de F\'{i}sica ``J. J. Giambiagi'' and IFIBA, FCEyN, Universidad de Buenos Aires, 1428 Buenos Aires, Argentina}

\author{Lucas C. C\'{e}leri}
\affiliation{Instituto de F\'{i}sica, Universidade Federal de Goi\'{a}s, 74001-970, Goi\^{a}nia,
Brazil}

\author{Norton G. de Almeida}
\affiliation{Instituto de F\'{i}sica, Universidade Federal de Goi\'{a}s, 74001-970, Goi\^{a}nia, Brazil}

\author{Augusto J. Roncaglia}
\affiliation{Departamento de F\'{i}sica ``J. J. Giambiagi'' and IFIBA, FCEyN, Universidad de Buenos Aires, 1428 Buenos Aires, Argentina}

\author{Fabricio Toscano}
\affiliation{Instituto de F\'{i}sica, Universidade Federal do Rio de Janeiro, 21941-972, Rio de Janeiro, Brazil}

\begin{abstract}

In the context of nonequilibrium quantum thermodynamics, variables like work behave stochastically. A particular definition of the 
work probability density function (pdf) for coherent quantum processes allows the verification of the quantum version of the celebrated 
fluctuation theorems, due to Jarzynski and Crooks, that apply when the system is driven away from an initial equilibrium thermal state. 
Such a particular pdf depends basically on the details of the initial and final Hamiltonians, on the temperature of the initial thermal state 
and on how some external parameter is changed during the coherent process. 
Using random matrix theory we derive a simple analytic expression that describes the 
general behavior  of the work characteristic function $G(u)$, associated with this particular work pdf for sudden quenches,
valid for all the traditional Gaussian ensembles of Hamiltonians matrices. 
This formula well describes the general behavior of $G(u)$ calculated from single draws 
of the initial and final Hamiltonians in all ranges of temperatures.

\end{abstract}
\maketitle

\section{Introduction}

The study of nonequilibrium thermodynamics in classical statistical mechanics has been mainly guided by the discovery of the 
so-called fluctuation theorems \cite{Bochkov1977,Evans1993,Gallavotti1995,Jarzynski1997,Crooks1999}. 
The quantum version of these theorems has recently been explored (see Refs. \cite{Esposito2009,HanggiReview,HanggiReviewErratum} for reviews). 
The theoretical and the experimental study of quantum fluctuation relations are of primary interest, both for fundamental issues and also for 
understanding the limitations for implementing new technological devices that work in the quantum regime \cite{HanggiReview,HanggiReviewErratum}. 
These devices are the basic components of the revolution that is unfolding in the field of quantum communication and information processing.

A fluctuation theorem relates the work performed on a given system  ---by a process that usually takes the system out of equilibrium--- with equilibrium properties. Classically, this work is described by a random variable and the quantities of interest must be averaged over an ensemble of phase space initial conditions. Therefore, such theorems rely on the full statistics of the work. The most common situation considered is when the process drives the system far away from an initial equilibrium thermal state. In the quantum realm the situation 
is a little bit more subtle due to the lack of an unambiguous definition of work. 
Remarkably, the important fluctuation theorems in Refs. \cite{Jarzynski1997,Crooks1999} have been extended to the quantum regime considering the two-projective-energy-measurements scheme \cite{Esposito2009,HanggiReview,HanggiReviewErratum,Kurchan2000,Tasaki2000,Talkner2007}. 
In this way, the statistics of the work has been investigated in
several contexts such as forced harmonic systems \cite{Talkner2008,Talkner2009,Deffner2008}, many-body systems \cite{Wang2017,halpern2017mbl},
and chaotic systems \cite{Zhu2016,Garcia-Mata2017-1,Garcia-Mata2017-2,Chenu2018}
among others.
 
In such a scheme, the system is initially prepared in a thermal state and then suffers a process that changes the Hamiltonian by a 
controlled parameter. In general the statistics of the quantum work depends on the specific characteristics of the 
time-dependent Hamiltonian that describes the process. Even in the cases of sudden quenches, where the unitary dynamics between the times 
of the initial and final Hamiltonians can be neglected \cite{Fusco2014}, the quantum work statistics depends on the characteristics of 
the initial and final Hamiltonians. Here we are interested in the description of such processes from the perspective of random matrix theory.

Random matrix theory (RMT) is applicable to systems where the Hamiltonians can be replaced by an ensemble of random Hamiltonians in
order to describe generic properties \cite{Guhr1997,Forrester2003}. The range of systems where RMT is successfully applied is broad, 
including the atomic nuclei of complex atoms \cite{Guhr1997,Weidenmuller2009}, other many-body systems \cite{Gubin2012,Forrester2003}, 
and quantum systems with classical chaotic counterparts \cite{Haake-book,Linda-book,Stockmann-book}. In such systems RMT successfully 
describes some spectral fluctuation properties, like the distribution of level spacing and many-level correlation functions.  

Regarding fluctuations theorems, it has been recently shown in Ref. \cite{Lobejko2017} that, for sudden quenches where the 
initial and final Hamiltonians belong to two independent Gaussian unitary ensembles (GUEs), the work probability distribution function
has a universal form that depends only on the general properties that define the ensembles. The authors also obtained analytical forms 
for the work probability distribution function in the limits of zero and infinite temperature.   

Here, using RMT, we have found an analytical expression for the ensemble average of the work characteristic function corresponding to the work
distribution of sudden quenches that is valid for an arbitrary temperature, considering all Gaussian ensembles. This result can help us to study 
the behavior of quantum chaotic systems regarding their thermodynamic properties. This is a very important issue since chaos and thermodynamics 
are deeply linked in classical mechanics while their connection in quantum systems is very cloudy. 

This paper is organized as follows. In Sec. \ref{SectionII} we briefly review the theory of nonequilibrium thermodynamics applied to quantum 
systems and discuss the necessary features of RMT necessary to develop our work, introducing the three possible Gaussian ensembles of random 
matrices and their spectral and eigenvectors probability densities. In Sec. \ref{SectionIII} we present our main result. 
We compare the obtained results with numerical simulations of random matrices taken from a Gaussian ensemble in Sec.\ref{Section-comparison}.  
Finally, in Sec. \ref{SectionV} we present our conclusions.

\section{Background}
\label{SectionII}

\subsection{Work statistics}

Let us consider a working process that changes the Hamiltonian of the system from $H:=H(0)$ to $\tilde H:=H(\tau)$, with $\tau$ being 
the duration of the process. We are interested in the statistics of the work performed by the external agent in order 
to implement such a process. To do this we consider the following protocol. A projective energy measurement $\Pi_{n}^{0}$ is performed on 
the system whose initial state is $\rho_{0}$, resulting in the energy eigenvalue $E_{n}$ with the probability 
$p_{n} = \mbox{Tr}\left[\rho_{0}\Pi_{n}^{0}\right]$.
After this, the process $U_{\tau}$ acts on the system and a second energy measurement $\Pi_{m}^{\tau}$ is performed, 
resulting in the eigenvalue $\tilde{E}_{m}$ with the probability 
$p_{m|n} = \mbox{Tr}\left[\Pi_{m}^{\tau}U_{\tau}\Pi_{n}^{0}\rho_{0}\Pi_{n}^{0}U_{\tau}^{\dagger}\right]/p_{n}$.
\par
The quantum work is then defined as $w:= \tilde{E}_{m} - E_{n}$, which is a random variables whose associated pdf is given by
\begin{equation}
P(w) = \sum_{m,n=1}^N p_{m,n}\delta\left[w - (\tilde{E}_{m} - E_{n})\right],
\label{defPw}
\end{equation}
where $p_{m,n} = p_{n}p_{m|n}$ is the joint probability density associated with the two energy measurements.
\par
With these definitions, the following quantum fluctuation relation \cite{Esposito2009,HanggiReview,HanggiReviewErratum} applies:
\begin{equation}
\left\langle e^{-\beta w}\right\rangle \equiv \int d w P(w) e^{-\beta w} = e^{-\beta \Delta F}.
\label{eq:fluc-rel}
\end{equation}
Here, we use $\Delta F := F_{\tau} - F_{0} = (-1/\beta)\log(\mathcal{Z}_{\tau}/\mathcal{Z}_{0})$ for the variation on the Helmholtz 
free energy, with $\mathcal{Z}_t=\mbox{Tr}e^{-\beta H(t)}$ being the partition function at time $t$ while $\beta = 1/T$ is the inverse temperature 
(Boltzmann constant is taken to unity). Therefore, we assumed that the initial state of the system is the thermodynamic equilibrium one
\begin{equation}
\rho_{0} = \frac{e^{-\beta H}}{\mathcal{Z}_0},
\label{eq:thermal}
\end{equation}
By defining the initial and final energy eigenbases as 
$\left\lbrace|\psi^{\gamma}_n\rangle\right\rbrace$ and $\left\lbrace|\tilde{\psi}^{\alpha}_{m}\rangle\right\rbrace$, respectively,
with $\gamma$ and $\alpha$ labeling a possible degeneracy,
we can rewrite 
the pdf associated with the quantum work as
\bea
P(w)&=& \sum_{m,n=1}^N \frac{e^{-\beta E_n}}{\mathcal{Z}_{0}} 
\sum_{\alpha,\gamma}\matrixel{\tilde \psi^{\alpha}_m}{U_{\tau}}{\psi^{\gamma}_n}
\matrixel{\psi^{\gamma}_n}{U^\dagger_{\tau}}{\tilde\psi^{\alpha}_m}\times \nonumber\\
&&\times\delta\left[w - (\tilde{E}_{m} - E_{n})\right],
\label{PWspecific}
\eea
where we have used that $\Pi_{n}^{0} = \sum_{\gamma}|\psi^{\gamma}_n \rangle \langle \psi^{\gamma}_n|$
and $\Pi_{m}^{\tau} = \sum_{\alpha}|\tilde{\psi}^{\alpha}_m \rangle \langle \tilde{\psi}^{\alpha}_m|$. That is,
$\Pi_{n}^{0}$ and $\Pi_{m}^{\tau}$ 
are the projectors onto the subspaces spanned by the eigenvectors associated with the possibly degenerated levels $E_{n}$ and $\tilde{E}_m$,
respectively.
\par  
Sometimes, it is easier to work with the characteristic function associated with the work pdf, which is defined as the Fourier transform 
of $P(w)$
\begin{equation}
G(u)=\int_{-\infty}^{\infty}P(w)e^{iuw}dw,
\label{chacfunc}
\end{equation}
which contains exactly the same information as $P(w)$. The main goal of this work is to provide, in the next section, 
the random matrix theory approach for the study of the characteristic function associated with sudden processes known as quantum quenches, i.e. 
processes that abruptly change some parameter of the Hamiltonian of the system, in such a way that $U_{\tau} \approx \mathds{1}$. 
In this case, using Eq.(\ref{PWspecific}), the characteristic function of the work becomes
\begin{equation}
G(u) = \sum\limits _{n,m=1}^N\frac{e^{-\beta E_n}}{\mathcal{Z}_0}\sum_{\alpha,\gamma}
|\braket{\tilde{\psi}^{\alpha}_{m}|\psi^{\gamma}_{n}}|^{2}
e^{iu(\tilde{E}_{m}-E_{n})}.
\label{chacfunc2}
\end{equation}

\subsection{Basic tools from RMT}
\label{SectionIIc}

The most common ensembles of the theory are the Gaussian ones. Such ensembles follow a classification according to their invariance properties
under time reversal symmetry $\Lambda$: The GUE, with no time-reversal invariance, 
the Gaussian orthogonal ensemble (GOE), and the Gaussian symplectic ensemble (GSE), corresponding to the two possibilities of time-reversal 
invariance, i.e., with $\Lambda^2 = 1$ (e.g., spin-even systems) or $\Lambda^2 = -1$ (e.g., spin-odd systems), respectively. 
The matrices representing such ensembles are $N\times N$ real symmetric (GOE), $N\times N$ Hermitian (GUE) and $2N\times 2N$ 
self-dual Hermitian matrices (GSE) \cite{Mehta-book}. One does need $\beta_e = 1, 2$ and $4$ parameters to define a matrix element
for GOE (real elements), GUE (complex elements) and GSE (quaternion real elements), respectively.

The statistical independence of the matrix elements in Gaussian ensembles of random matrices, $\mathbb H$, forces the joint 
probability density to be of the form $P({\bf H}) \propto e^{-\frac{1}{2\sigma^2}\;[\Tr(\mathbb H^{2})-2\expval{E}\Tr(\mathbb H)]}$, 
where ${\bf H}$ is the vector of matrix elements of $\mathbb H$. The parameters $\expval{E}$ and $\sigma^2$ are, respectively, the mean value 
of the diagonal elements and the variances of the diagonal as well as of the real and 
imaginary parts of the off-diagonal elements \cite{Mehta-book}. While $\expval{E}$ fixes the center of the average 
random matrix spectrum, $\sigma$ sets the energy scale.  

Here we do not explicitly show the different measures $d{\bf H}$ for the three Gaussian ensembles in the space of 
Hermitian matrices \cite{Mehta-book,Linda-book}. Instead, we express the joint probability density $P({\bf H})$ in the polar form 
using the invariance of the ensembles under unitary transformations parametrized by the set of angles 
$\boldsymbol{\theta}=(\theta_{1},\ldots,\theta_{\beta_eN(N-1)/2})$. Therefore, 
by defining the vector of eigenvalues ${\bf E}\equiv (E_1,\ldots,E_N)$ and considering that $\Tr({\mathbb H})=a_{\beta_e}\sum_{j=1}^N E_j$
and $\Tr({\mathbb H}^2)=a_{\beta_e}\sum_{j=1}^N E_j^2$, with $a_{\beta_e}=1$ for $\beta_e=1$ and $2$ and $a_{\beta_e}=2$ for $\beta_e=4$, 
the joint probability density can be written as \cite{Linda-book}
\beq
\label{joint-prob-E-theta}
P({\bf E},\boldsymbol{\theta})\propto e^{-\frac{a_{\beta_e}}{2\sigma^2}\left[\sum_{j=1}^NE_j^2-2\expval{E}\sum_{j=1}^N E_j\right]}J({\bf E},\boldsymbol{\theta}),
\eeq
where $J({\bf E},\boldsymbol{\theta}) =\left|\partial {\bf H}/\partial ({\bf E},\boldsymbol{\theta})\right|=|\varDelta_{N}({\bf E})|^{\beta_{e}}
P({\boldsymbol{\theta}})$ is the Jacobian, $\varDelta_{N}({\bf E})= \prod_{1\leq j<k\leq N}(E_{k}-E_{j})$ is the Vandermonde 
determinant and $P$ is some function of $\boldsymbol{\theta}$. The integration over $\boldsymbol{\theta}$ gives the joint probability 
density for the eigenvalues
\beq
\label{eigendist}
P({\bf E})=  \tilde C_{N,\beta_e}\;e^{-\frac{a_{\beta_e}}{2\sigma^2}\sum\limits _{k=1}^{N}(E_{k} -  \langle E \rangle)^{2}}|
\varDelta_{N}({\bf E})|^{\beta_{e}},
\eeq
with $\langle E \rangle=(1/a_{\beta_e}N)\int d{\bf E}\,P({\bf E})\,E_j$.  The formal expression for the normalization 
constant $\tilde C_{N,\beta_e}$ will not be necessary in this study. We refer to Ref. \cite{Mehta-book} for details.

The integration of the joint probability density in Eq. (\ref{joint-prob-E-theta}) over ${\bf E}$ gives $P(\boldsymbol{\theta})$, that is, 
the joint pdf for the complete set of eigenvectors. One can clearly see the statistical independence of the eigenvectors and 
eigenenergies, $P({\bf E},\boldsymbol{\theta})=P({\bf E})P(\boldsymbol{\theta})$, 
with $\int d{\bf E}\;P({\bf E})\int d\boldsymbol{\theta} \; P(\boldsymbol{\theta})=1$. Although $P(\boldsymbol{\theta})$ is associated 
with the set of eigenvectors $\{{\bf v}_j\}_{j=1,\ldots,a_{\beta_e}N}$ \footnote{With the components being real numbers for the GOE ensemble
and complex numbers for the GUE and GSE ensembles.} it is more useful to have a joint probability density that is a function of the 
components of each one of the vectors ${\bf v}_j$: $P(\{{\bf v}_j\}_{1,\ldots,a_{\beta_e}N})$. However, the imposition of the orthonormal 
condition over the eigenvectors on this density function is a cumbersome task \cite{Linda-book}. Instead, what one usually does 
is to consider the probability density of a single eigenvector taken from the complete set $P({\bf v}_j)$. This probability density is
considered uniform in an hypersphere defined by the components of the eigenvectors. In the limit of large dimension $N$ these components 
become random independent variables Gaussianly distributed. In general, the statistical properties associated with the eigenvectors of 
Gaussian ensembles are calculated using this approximation. 
\par
We can use the Gaussian approximation for the distribution of single eigenvectors to calculate the average 
$\expval*{| \ip*{\tilde{\psi}^{\alpha}_m}{\psi^{\gamma}_n}|^2}_{vec}$
that will be useful in the next section. We consider two Hamiltonians  
with completely uncorrelated energy spectra and with sets 
of eigenstates given by $\{| \psi^{\gamma}_n \rangle \}$ and $\{| \tilde{\psi}^{\alpha}_m \rangle \}$, $n,m =1,\ldots,N$ and 
$\alpha, \gamma = 1,..., a_{\beta_e}$.
The two Hamiltonians are drawn from different Gaussian ensembles of the same kind, characterized by the sets of 
parameters $\{\expval{E},\sigma\}$ and $\{\expval*{\tilde E},\tilde\sigma\}$. We denote by $\expval{\ldots}_{vec}$ the mean value with 
respect to the independent densities $P({\bf v}_n)$ and $P(\tilde{\bf v}_m)$ with 
$({\bf v}_n)_j:=\ip{\phi_j}{\psi^{\gamma}_n}$ and $(\tilde{\bf v}_m)_j:=\ip*{\phi_j}{\tilde{\psi}^{\alpha}_m}$, 
($\{|\phi_j \rangle \}$ with $j =1,\ldots, a_{\beta_e}N$ being a complete basis of the Hilbert space).
Therefore we have
\bea
\expval*{| \ip*{\tilde{\psi}^{\alpha}_m}{\psi^{\gamma}_n}|^2}_{vec} &=& \sum_{j, k = 1}^{a_{\beta_e} N}
\expval{(\tilde{\bf v}_m)^*_j(\tilde{\bf v}_m)_k}
\expval{({\bf v}_n)_j({\bf v}_n)^*_k} \approx \nonumber \\
&\approx& \frac{1}{a_{\beta_e}N}, 
\label{mediaRMTvec}
\eea
where we used that $\expval{({\bf v}_n)_j({\bf v}_n)^*_k}\approx
1/(a_{\beta_e}N)\delta_{jk}$ (and $\expval{(\tilde{\bf v}_m)^*_j(\tilde{\bf v}_m)_k}\approx 1/(a_{\beta_e}N)\delta_{jk}$). 
The result in Eq.(\ref{mediaRMTvec}) shows that there are no privileged directions in Hilbert space associated with the Gaussian ensembles.
\par
Central to our study is the behavior of the level density for large spectra belonging to the Gaussian ensembles described before, which is 
defined by
\beq
\rho(E) = N\int d{\bf E} \;P({\bf E})\;\delta(E_j-E) = N\expval{\delta(E_j-E)}.
\label{dens}
\eeq
Note that $\int_{-\infty}^{\infty} dE\;\rho(E)=N$. For $N\gg 1$ one can prove that the level density for the three ensembles has the 
same asymptotic behavior, given by the ``semicircle law'' formula \cite{Mehta-book}:
\beq
\label{sc-law}
\rho_{N\gg 1}(a,x)= \left\{ \begin{matrix} 
      \frac{2N}{\pi a}\sqrt{1-\left(\frac{x}{a}\right)^2}&, \mbox{
    $\frac{|x|}{a} \le 1$} \\
      0 &,  \mbox{
    $\frac{|x|}{a} > 1$} \\
   \end{matrix}\right.,
\eeq
with $x$ and $a$ real numbers ($a>0$). For Gaussian ensembles we have  $x=E-\expval{E}$  and  $a=\sqrt{2N\beta_e}\sigma$. 
An important statistical parameter that can be defined from the level density in 
Eq. (\ref{sc-law}) is the average level spacing at the center of the spectrum:
\begin{equation}
\label{mediumspace}
\expval{s} = \frac{1}{ \rho_{N\gg 1} (a,0)}=
\pi \sigma\sqrt{\frac{\beta_e}{2N}}\,,
\end{equation} 
where $s = E_{j+1}-E_j$, and therefore for Gaussian ensembles we can rewrite the positive parameter in Eq.(\ref{sc-law}) as
$a = 2N\expval{s}/\pi$.
Then, instead of using the set of numbers $\{N,\beta_e,\expval{E}, \sigma \}$ to define 
a Gaussian ensemble we can alternatively use the set of numbers $\{N,\beta_e,\expval{E}, \expval{s}\}$.  So, the level 
density for $N\gg 1$ is given by $\rho_{N\gg 1}(2N\expval{s}/\pi, E-\expval{E})$, i.e., the same function for the three ensembles.
\par
An important feature of Gaussian ensembles is that they have a sort of ergodic property: For sufficiently 
large dimension $N$ the running average of some spectral measure, calculated over the spectrum of a single random matrix, 
is approximately equal to the ensemble average \cite{Weidenmuller2009}. This property 
will be used to tackle the problem 
of finding a general behavior for the characteristic function of the work of Gaussian random matrices subjected to sudden quenches.

\section{RMT approach to the characteristic function of work}
\label{SectionIII}

Here we  consider RMT to describe the probability density of the work associated with a sudden quench connecting two Hamiltonians 
taken from one of the three Gaussian ensembles described in Sec. \ref{SectionIIc}. We do this in the limit of 
large dimension, viz., $ N\gg 1$, in which the level density of the spectra is given by the semicircle law in Eq.(\ref{sc-law}). 
We denote the spectrum of the initial Hamiltonian (before the quench) by ${\bf E}$ and the respective set 
of eigenstates by $\{| \psi^{\gamma}_j \rangle \}$, with $j=1,\ldots,N$ and $\gamma = 1,\ldots,a_{\beta_e}$. Likewise, the spectrum
of the final Hamiltonian (after the quench) 
is represented by $\tilde {\bf E}$, whose associated eigenstates are $\{| \tilde{\psi}^{\alpha}_j \rangle \}$, with
$\alpha = 1,\ldots,a_{\beta_e}$.
The sets of parameters 
that characterize the spectra are $\{\expval{s},\expval{E}\}$ and $\{\expval{\tilde s},\langle{ \tilde E}\rangle\}$ [see Eq. (\ref{mediumspace})].
Therefore, the quench is given by the process 
$\{\expval{s},\expval{E}\} \rightarrow \{\expval{\tilde s},\langle{ \tilde E}\rangle\}$.
\par
We begin by computing the average of Eq.\eqref{chacfunc2} over the Gaussian random matrix probability densities of eigenvalues
and eigenvectors, 
described in the last section, noticing the independence between them. This procedure gives us
\begin{eqnarray}
&\expval{G(u)} = 
\sum\limits _{n,m=1}^{N} \expval{\frac{e^{-\beta E_n}}{\mathcal{Z}_0} e^{-iuE_n}}
\left< e^{iu\tilde{E}_m}\right>
 \nonumber\\
&\times \sum\limits_{\alpha,\gamma = 1}^{a_{\beta_e}}\expval*{|\braket{\tilde{\psi}^{\alpha}_{m}|\psi^{\gamma}_{n}}|^{2}}_{vec}. 
\label{GuRMTfirst}
\end{eqnarray}
Here the brackets $\left<...\right>$ and  $\left<...\right>_{vec}$ stand for the average over the eigenvalues and over the eigenvectors, 
respectively, and we used the fact that the spectra ${\bf E}$ and $\tilde{\bf E}$ are completely 
uncorrelated \footnote{The absence of correlations between the initial and final spectra of the Hamiltonians is a typical characteristic of 
very strong quenches.}. We also considered that the initial and final eigenvalues, $E_n$ and $\tilde E_m$, are statistically 
equivalent and that there are no privileged directions in the Hilbert spaces spanned by 
the basis $\{\ket{\psi_n}\}$ and $\{\ket{\tilde \psi_m}\}$, as we already mentioned in the previous section. 
This means that neither of the averages in Eq.(\ref{GuRMTfirst}) 
depend on the indexes $m$ and $n$. Therefore,
\beq
\expval{G(u)} = 
a_{\beta_e}^2 N^2 
\frac{\expval{e^{-\beta E}e^{-iuE}}}
{\expval{\mathcal{Z}_0}} 
\expval{ e^{iu\tilde{E}}}
\expval*{ |\braket{\tilde{\psi}|\psi}|^{2}}_{vec}.
\label{N2terms}
\eeq
Here, we further use the approximation
\beq
\expval{\frac{e^{-\beta E}}{\mathcal{Z}_0} e^{-iuE}} \approx \frac{\expval{e^{-\beta E}e^{-iuE}}}{\expval{{\cal Z}_0}}=\frac{\expval{e^{-\beta E}e^{-iuE}}}{a_{\beta_e}N\expval{e^{-\beta E}}},
\label{approxcrucial}
\eeq
which is compatible with the condition $G(0)=1$, which stems from the definition of the characteristic function 
in Eq.(\ref{chacfunc}), because within the scope of this approximation we have
$\expval{G(0)}=1$.
In Sec. \ref{Section-comparison} we discuss the range of validity of the approximation in Eq.(\ref{approxcrucial}). 
\par
We must now perform the averages given in the previous equations. For large values of $N$ we have
\bse
\label{eqmultiples}
\bea
\expval{e^{iu\tilde{E}}}&\approx&
\frac{1}{N}\int_{-\infty}^{\infty}d{\tilde E}\;e^{ iu{\tilde E}}\, \rho_{N\gg 1}(\tilde E)\nonumber\\
&=&2\;e^{iu\langle\tilde{E}\rangle}\;\frac{J_{1}(2N \expval{\tilde s} u/\pi)}{2N \expval{\tilde s} u/\pi}\;,
\label{finaltext}
\\
\label{avpartition}
\expval{e^{-\beta E}}  
&\approx&\frac{1}{N}\int_{-\infty}^{\infty}d{E}\;e^{-\beta E}\, \rho_{N\gg 1}(E) \nonumber\\
&=& 2\;e^{-\beta \langle E \rangle}\;\frac{I_{1}(2N \expval{ s} \beta/\pi)}{2N \expval{ s} \beta/\pi},\\
\expval{e^{-\beta E}e^{-iuE}}&\approx&
\frac{1}{N}\int_{-\infty}^{\infty}d{E}\;e^{-\beta E}e^{-iuE}\, \rho_{N\gg 1}(E)=\nonumber\\
&=& e^{-\beta \langle E \rangle}e^{-iu\langle E \rangle}\,\times\nonumber\\
&\times& _{0}F_{1}\left[2\;, -N^2\expval{s}^2(u - i\beta)^2/\pi^2 \right],
\label{avpartitionbeta}
\eea
\ese
with $J_{n}\left(x\right)$ and $I_n(x)$ being the Bessel and modified Bessel functions of the first kind, respectively, 
while $_{0}F_{1}\left[c,x\right]$ stands for the confluent hypergeometric function. The details are given in the Appendix.

Gathering the results in Eqs.(\ref{eqmultiples}) and (\ref{mediaRMTvec}) into Eq.(\ref{N2terms}), we finally get
\begin{eqnarray}
&\expval{G(u)} \approx \;
e^{iu (\langle\tilde{E}\rangle-\langle E\rangle) }\,
\frac{2N \expval{s}\beta/\pi}
{I_{1}(2N \expval{s}\beta/\pi)}\nonumber\\
&\times\,_{0}F_{1}\left[2\;, -N^2\expval{s}^2(u - i\beta)^2/\pi^2 \right]\,\frac{J_{1}(2N \expval{\tilde s} u/\pi)}{2N \expval{\tilde s} u/\pi}.
\label{GuRMTgeral}
\end{eqnarray}
Using the identity $_{0}F_{1}[2,x^2]=I_1[2x]/x$ 
and $\lim_{x\rightarrow 0} J_1(x)/x=1/2$, we can easily check that the expression in Eq.(\ref{GuRMTgeral}) satisfies $\expval{G(0)}=1$.
 
This is the main result of our work. In Sec. \ref{Section-comparison} we see that Eq.(\ref{GuRMTgeral}) essentially reproduces, 
in the regime $N\gg 1$ and for all values of the inverse temperature $\beta$, the behavior of the characteristic function $G(u)$ calculated 
from single members, $H$ and $\tilde H$, of all Gaussian ensembles. But first let us analyze 
two important limits: $\beta\rightarrow 0$ and 
$\beta \rightarrow +\infty$.

\subsection{Infinite temperature ($\beta=0$)}

In order to compute the limit $\beta \rightarrow 0$ in Eq.(\ref{GuRMTgeral}) we use the fact that, for real values of $x$,
we have $\lim_{x\rightarrow 0} I_1(a\,x)/(a\, x)=1/2$ and 
$_{0}F_{1}[2,-x^2]=J_1[2x]/x$, thus resulting in
\beq
\expval{G(u)}_{\beta=0}=e^{iu (\langle\tilde{E}\rangle-\langle E\rangle) }\,
\frac{J_{1}(2N \expval{ s} u/\pi)}{2N \expval{s} u/\pi}.
\frac{J_{1}(2N \expval{\tilde s} u/\pi)}{2N \expval{\tilde s} u/\pi}.
\label{medGu0}
\eeq
Note that Eq. (\ref{medGu0}) is the result that we would obtain by averaging the exact expression of 
the characteristic function when $\beta=0$ [the limit $\beta\rightarrow 0$ in Eq. (\ref{chacfunc2})], viz.,
\beq
G_{\beta=0}(u)=\sum\limits _{n,m=1}^N\frac{1}{a_{\beta_e}N}
\sum\limits_{\alpha,\gamma = 1}^{a_{\beta_e}}|\braket{\tilde{\psi}^{\alpha}_{m}|\psi^{\gamma}_{n}}|^{2}
e^{iu(\tilde{E}_{m}-E_{n})}.
\label{Guexactbetazero}
\eeq
Then, using the approximations given by Eqs. (\ref{mediaRMTvec}) and (\ref{finaltext}) and that
\beq
\label{medEumenos}
\expval{e^{-iuE}}\approx 2e^{-iu\expval{E}}\frac{J_1(2N\expval{s}u/\pi)}{2N\expval{s}u/\pi},
\eeq
which is the analogous to that in Eq. (\ref{finaltext}), we can perform the ensemble average over $G_{\beta=0}(u)$ in 
Eq.(\ref{Guexactbetazero}), obtaining exactly the same expression of Eq.(\ref{medGu0}). This happens because both 
sides of Eq.(\ref{approxcrucial}) are equal and, in the regime $N\gg 1$, the approximations in Eqs. (\ref{avpartition}) and 
(\ref{avpartitionbeta}) are extremely accurate for all values of $\beta$.

From the inverse Fourier transform of Eq. (\ref{medGu0}) we recover the probability density of the work, within RMT, for infinite temperature, 
which corresponds to the convolution of two semicircle law functions:
\bea
&\expval{P(w)}_{\beta=0}=\frac{1}{4N^2}\int du \,\rho_{N\gg 1}\left(\frac{2N\expval{s}}{\pi},u\right)\nonumber\\
&\times\rho_{N\gg 1}\left(\frac{2N\expval{\tilde s}}{\pi},w-(\langle\tilde E\rangle-\expval{E})-u\right).
\eea 
This result was essentially derived in Ref. \cite{Lobejko2017} in the context of the GUE but here we show that it is valid for 
all Gaussian ensembles providing $N\gg 1$. Remembering that the semicircle law function, $\rho_{n\gg 1}(a,x)$, is an even function 
with compact support, which has a maximum at the origin $x=0$ and a width of $2a$, we can estimate that $\expval{P(w)}_{\beta=0}$ has a 
peak and width given by
\bse 
\label{peakandwidth1}
\bea
w_*&=&\langle\tilde E\rangle-\expval{E},
\label{w*1}\\
\delta w &=& (2N/\pi)\max\{\expval{s},\expval{\tilde s}\}+\langle\tilde E\rangle-\expval{E},
\label{deltaw1}
\eea
\ese
respectively.
For $w>\delta w$ we have $\expval{P(w)}_{\beta=0}=0$.

\subsection{Zero temperature ($\beta\rightarrow +\infty$)}
\label{Section4b}

We can calculate the limit $\beta\rightarrow +\infty$ in Eq.(\ref{GuRMTgeral}) by noting that 
\bea
&\frac{2N\expval{s}\beta/\pi}{I_1(2N\expval{s}\beta/\pi)}\,_{0}F_{1}\left[2,-\frac{N^2\expval{s}^2}{\pi^2}(u - i\beta)^2\right]
\nonumber\\
&\approx2\left(1+i\frac{u}{\beta}\right)^{-3/2}(e^{i\frac{2N\expval{s}}{\pi}}+e^{i\frac{2N\expval{s}}{\pi}}e^{-\frac{4\beta N\expval{s}}{\pi}})\nonumber\\
&\stackrel[\beta\rightarrow +\infty]{\approx}{}2\,e^{i\frac{2N\expval{s}}{\pi}},
\label{app2fase}
\eea
where we have used the approximations
\begin{equation*}
_{0}F_{1}[2,z]=(1/2\sqrt{\pi})(z)^{-3/4}(e^{2\sqrt{z}}+ie^{-2\sqrt{z}})+{\cal O}(1/z)^{5/4}
\end{equation*}
(where $\sqrt{z}$ is the square root with a positive real part of the complex number $z$) and 
\begin{equation*}
\frac{z}{I_1(z)}=\frac{e^z}{(-i+e^{2z})(1/z)^{3/2}+{\cal O}(1/z)^{5/2}},
\end{equation*}
which are accurate for large values of $|z|$. Therefore, by substituting Eq.(\ref{app2fase}) into Eq. (\ref{GuRMTgeral}), we obtain 
\bea
\expval{G(u)}_{\beta\rightarrow +\infty}&=&2\,
e^{iu\left(\langle\tilde{E}\rangle-\left(\langle E\rangle -\frac{2N\expval{s}}{\pi}\right)\right)}
\nonumber\\
&&\times\frac{J_{1}(2N \expval{\tilde s} u/\pi)}{2N \expval{\tilde s} u/\pi}.
\label{medGinfinito}
\eea
Instead, we can compute $\expval{G(u)}$, when $\beta \rightarrow +\infty$, if we directly perform the ensemble average over the exact expression,
\beq
G_{\beta\rightarrow +\infty}(u)=\frac{e^{-iuE_1}}{a_{\beta_e}}\sum\limits _{m=1}^N
\sum\limits_{\alpha,\gamma = 1}^{a_{\beta_e}}|\braket{\tilde{\psi}^{\alpha}_{m}|\psi^{\gamma}_{1}}|^{2}
e^{iu(\tilde{E}_{m})},
\label{Guliminfty}
\eeq
obtained by noting that $\lim_{\beta\rightarrow +\infty}e^{-\beta E_n}/{\cal Z}=\delta_{n,1}/a_{\beta_e}$ in Eq. (\ref{chacfunc2}). 
Using the 
results in Eqs. (\ref{mediaRMTvec}) and (\ref{finaltext}), and considering that now the ground-state energy, $E_1$, is a non fluctuating 
variable, we get  
\bea
\expval{G(u)_{\beta\rightarrow +\infty}}&\approx& 2\,e^{iu(\langle\tilde{E}\rangle-E_1) } \frac{J_{1}(2N \expval{\tilde s} u/\pi)}{2N \expval{\tilde s} u/\pi}.
\label{Gutempzeroboa}
\eea
When $N\gg 1$, we have $E_1\approx \langle E\rangle -2N\expval{s}/\pi$, leading to 
$\expval{G(u)_{\beta\rightarrow +\infty}}\approx\expval{G(u)}_{\beta\rightarrow +\infty}$.

From the inverse Fourier transform of Eq. (\ref{medGinfinito}) we recover the work pdf, within RMT, at zero temperature, as
\beq
\expval{P(w)}_{\beta\rightarrow +\infty}=\frac{1}{N}\rho_{N\gg 1}
\left(\frac{2N\expval{\tilde s}}{\pi},w-\left(\langle\tilde E\rangle-E_1\right)\right),
\label{Pwsemicirclelaw}
\eeq
where we have used that $E_1\approx \langle E\rangle -2N\expval{s}/\pi$.
Therefore, in the limit of zero temperature the probability density $\expval{P(w)}$ is proportional to a semicircle law function 
centered at $w_*$ and with a width $\delta w$ given by 
\bse
\label{peakandwidth2}
\bea
w_*&=&\langle\tilde E\rangle-E_1,\\
\delta w &=&2N\expval{\tilde s}/\pi.
\label{deltaw2}
\eea
\ese
The result in Eq. (\ref{Pwsemicirclelaw}) was essentially derived in Ref. \cite{Lobejko2017} in the context of the GUE, but here we show 
that it is valid for all Gaussian ensembles providing $N\gg 1$.
 
\section{Comparisons with numerical calculations and analysis}
\label{Section-comparison}

\begin{figure}[h!]
\centering 
\includegraphics[scale=0.5]{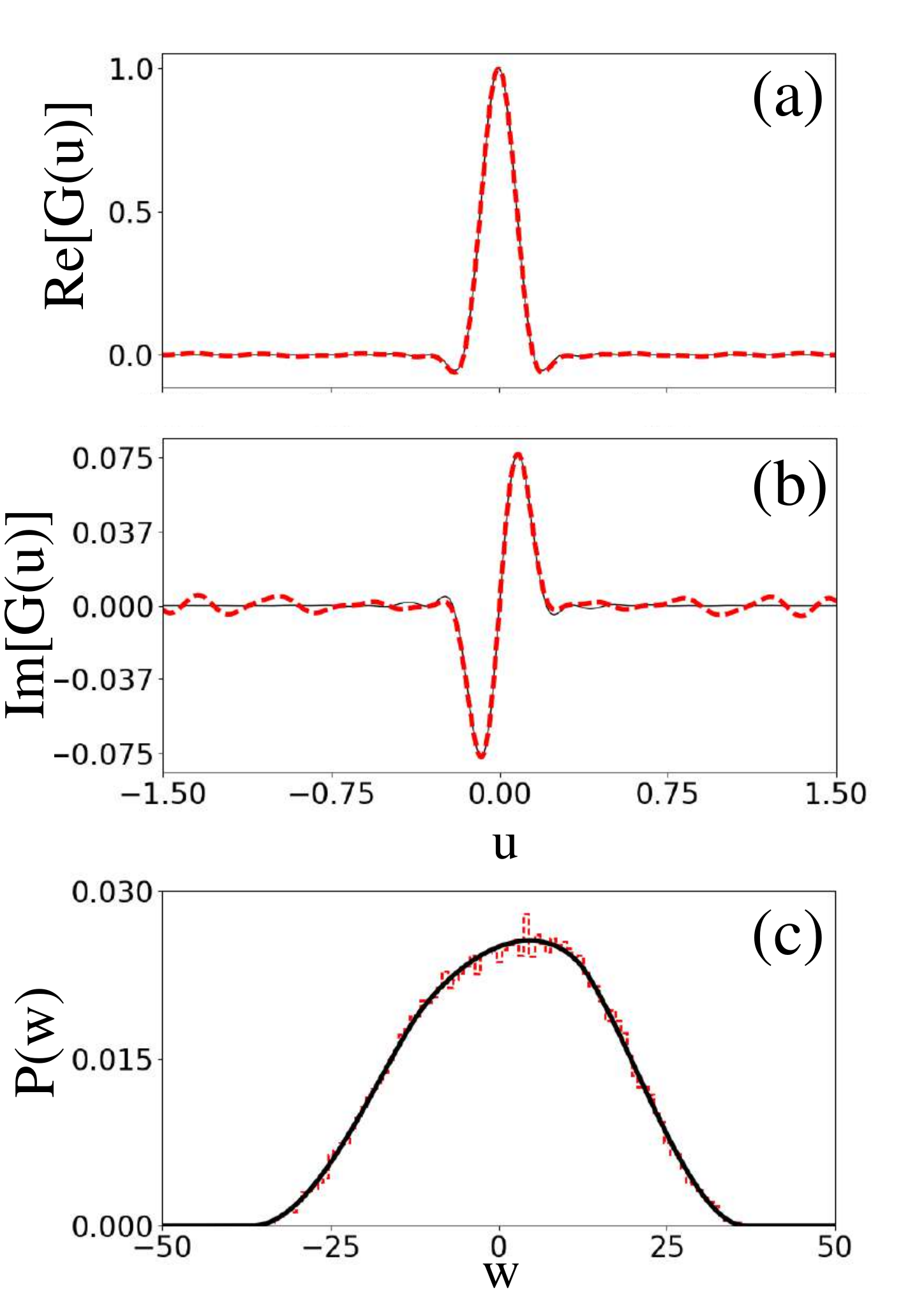}
\caption{(color online) In panels  {\bf (a)} and {\bf (b)} we plot the real and imaginary part of the work characteristic 
functions given in Eq. (\ref{GuRMTgeral}) (solid line) and calculated from Eq. (\ref{chacfunc2}) using single draws of the initial, 
$H$, and final, $\tilde H$, Hamiltonians of two GOEs (dashed lines). The parameters that define the ensembles 
are $N=300$, $\expval{s}=0.1283$, and $\expval{\tilde s}=0.1283/2$, and we chose $\expval{E}=\langle\tilde E\rangle\approx 24$
(arbitrary units of energy) 
such with this choice we displace both spectra in order to have $E_1=0$. 
The inverse temperature is $\beta=0.01$ so $N_{\rm eff}/N \approx 2.6$. 
In panel {\bf (c)} we plot the work probability density, as an histogram function (dashed line) calculated from the definition
in  Eq.~\eqref{defPw} and from the Fourier transform of Eq.~\eqref{GuRMTgeral} (solid line).   See text for details.}
\label{fig1} 
\end{figure}

Here we compare the expression of the work characteristic function given in Eq.(\ref{GuRMTgeral}) and the exact expression of this 
function given in Eq.(\ref{chacfunc2}), when it is calculated from single draws of the initial and final Hamiltonians of the GOE. We did 
this comparison only for the GOE because $\expval{G(u)}$ in Eq.(\ref{GuRMTgeral}) is the same for all three Gaussian ensembles. The analysis
was performed for three values of the inverse temperature that essentially comprise the behavior of the work characteristic function from 
very low to very large values of the temperature. We characterize low or high temperatures by defining $N_{\rm eff}:=1/(\beta\expval{s})$, 
a real number that is proportional to the effective number of energy levels of the initial Hamiltonian, $H$, that contribute to build the work 
characteristic function \footnote{$N_{\rm eff}$ is defined such that $e^{-\beta N_{\rm eff}\expval{s}}= 1/e$.}.Then, for a fixed value of mean 
level spacing 
for the initial spectrum, $\expval{s}$, and a fixed dimension of the ensemble, $N$,  low temperatures
correspond to $N_{\rm eff}/N \ll 1$ and high temperatures correspond to  $N_{\rm eff}/N> 1$ meaning that all the levels contribute to 
the work characteristic function. 
\par
We used two ensembles of Gaussian orthogonal matrices of dimension $N=300$. The mean level spacings that characterize the initial and final 
spectra are $\expval{s}=0.1283$ and $\langle \tilde s \rangle=0.1283/2$, respectively. First we draw the matrices $H$ and $\tilde H$ from 
the ensembles using $\expval{E}=\langle\tilde E\rangle=0$. Then, we displace the two spectra of eigenvalues by 
the quantity $\expval{E}=\langle\tilde E\rangle\approx 24$ (arbitrary units), which guarantees that the ground state energy of the initial
spectrum is zero. This procedure, of course, does not affect the eigenvectors.
 
\begin{figure}[h!]
\centering 
\includegraphics[scale=0.5]{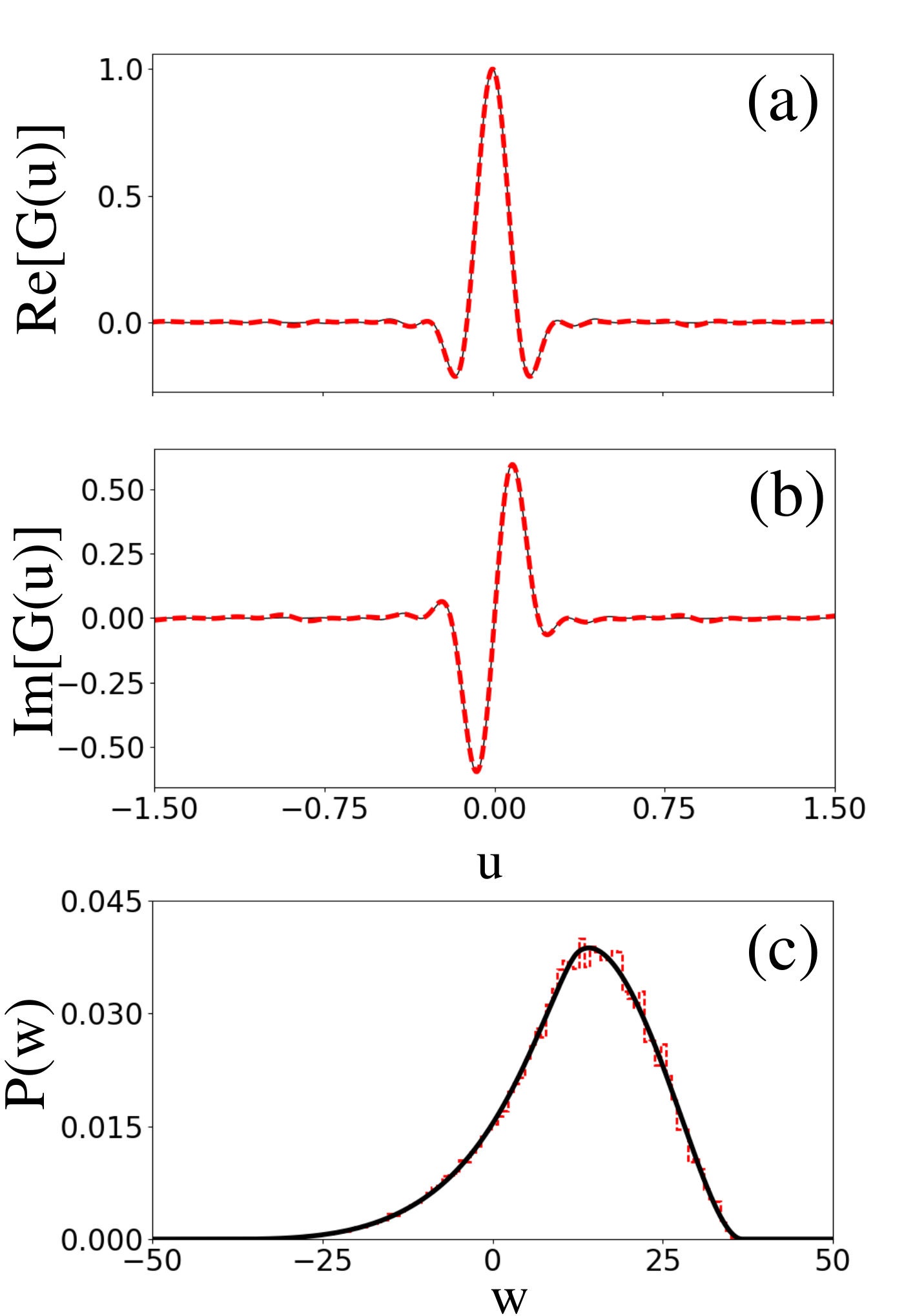}
\caption{(color online) The same as Fig. \ref{fig1} but with $\beta=0.1$.
Therefore $N_{\rm eff}/N\approx 0.26$ (see text for details)}. 
\label{fig2} 
\end{figure}

\begin{figure}[h!]
\centering 
\includegraphics[scale=0.5]{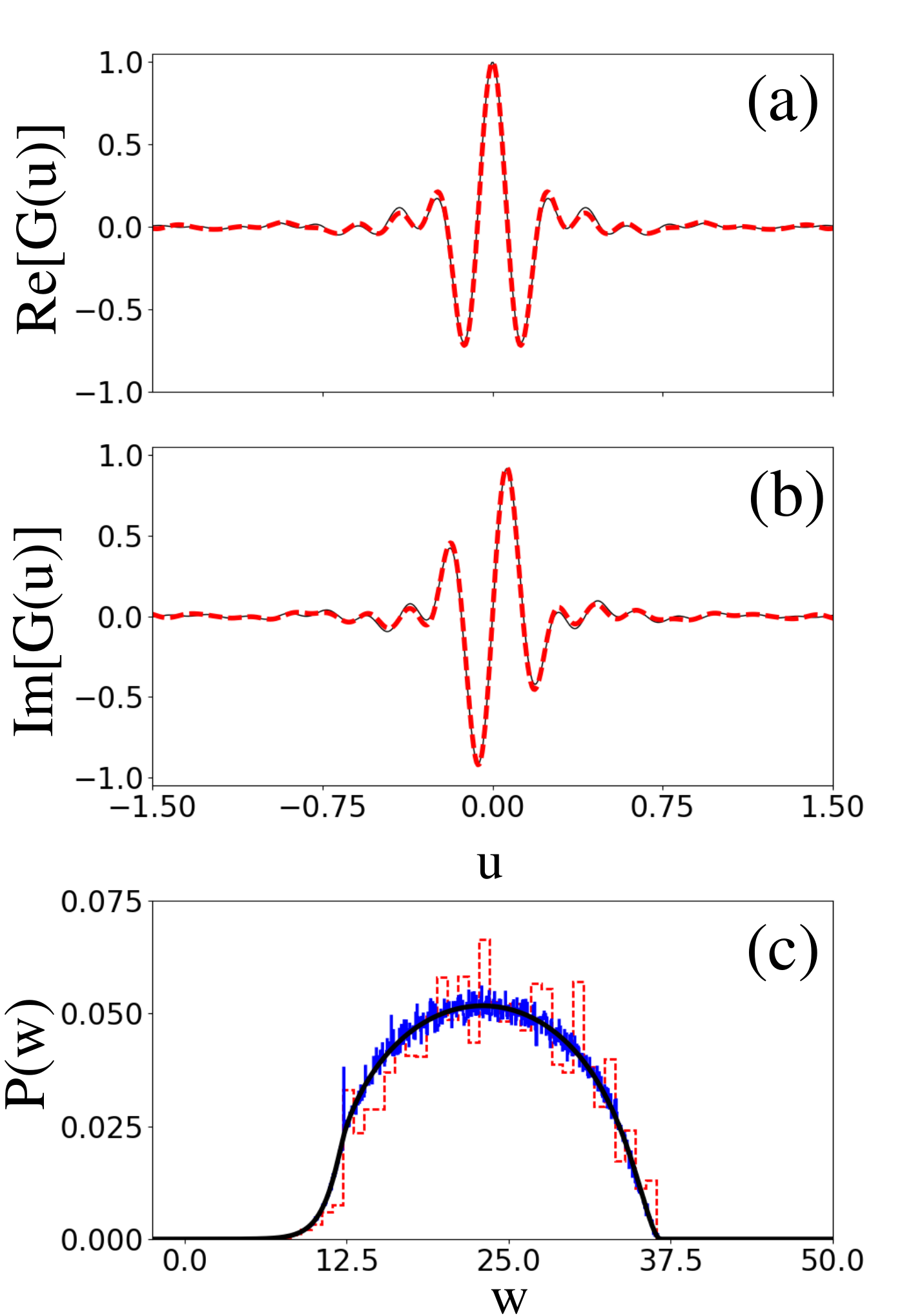}
\caption{(color online) The same as Fig.\ref{fig1} but with $\beta=1$ corresponding to
$N_{\rm eff}/N\approx 0.026$, in panels (a) and (b) and the dashed curve in panel (c). The thin solid line in panel (c) corresponds 
to $\expval{P(w)}$ calculated from two GOEs 
with $N=5000$ but keeping the values $\beta =1$ and  $N_{\rm eff}/N\approx 0.026$  (see text for details).}  
\label{fig3} 
\end{figure}

Figure \ref{fig1} shows an example for high temperature with $N_{\rm eff}\approx 2.6 > 1$. Both the real [Fig.\ref{fig1}{(a)}] 
and imaginary [Fig.\ref{fig1}{(b)}] parts of $\expval{G(u)}$ are in good agreement with the real and imaginary parts of 
the work characteristic function calculated from single draws using Eq.~\eqref{chacfunc2}. The same results can be obtained for 
an intermediate value of the temperature, as shown in Fig. \ref{fig2} with ${N_{\rm eff}/N\approx 0.26<1}$, and in Fig. \ref{fig3} for a
lower value of the temperature with ${N_{\rm eff}/N\approx 0.026}$ [except the thin solid curve in panel (c)].

We display the comparison between the work probability densities in panels {\bf (c)} of all figures. 
In all cases $\expval{P(w)}$, calculated from the inverse Fourier transform of $\expval{G(u)}$ in Eq.(\ref{GuRMTgeral}), 
describes the essential behavior of $P(w)$. Where $P(w)$ is obtained  
from the definition in Eq.~(\ref{defPw}), with $p_n=e^{-\beta E_n}/\mathcal{Z}_0$ and $p_{m|n}=|\braket{\tilde{\psi}^{\alpha}_{m}|\psi^{\gamma}_{n}}|^{2}$,
using the eigenvalues and eigenvectors of the matrices $H$ and $\tilde H$ from single draws of two GOEs. 
The ergodic property of the Gaussian ensembles implies that in the limit 
${N\rightarrow +\infty}$ the running average calculated from single draws
of two GOEs using Eq.~\eqref{defPw} coincides with $\expval{P(w)}$, the distribution
calculated from the inverse Fourier transform of $\expval{G(u)}$ in Eq.~(\ref{GuRMTgeral}).
In the examples presented in Figs.~\ref{fig1} and \ref{fig2}, for high and intermediate temperatures respectively, the matrix dimension 
$N=300$ is big enough to see the ergodic property working. However, for low temperature we considered two different dimensions,
$N=300$ and $N=5000$, 
to show that the running average of the probability distribution of the work converges to $\expval{P(w)}$. 
Therefore, in Fig.~\ref{fig3}(c) we fixed  
the low-temperature condition $N_{eff}/N\approx 0.026$ corresponding to 
 $\beta=1$ in two different initial ensembles characterized by the parameters 
 $\{N=300,\beta_e=1,\expval{E}\approx 24, \expval{s}=0.1283\}$ and 
 $\{N=5000,\beta_e=1,\expval{E}\approx 24, \expval{s}=0.1283\times 300/5000\}$. The corresponding final ensembles are 
 characterized by the parameters 
 $\{N=300,\beta_e=1,\langle \tilde E\rangle\approx 24, \langle \tilde s\rangle=0.1283/2\}$ and 
 $\{N=5000,\beta_e=1,\langle \tilde E\rangle\approx 24, \langle \tilde s\rangle=(0.1283/2)\times 300/5000\}$, where both have the same 
 value of $N\langle \tilde s\rangle$.
Notice that by fixing the values of $N\expval{s}$ and $N\langle \tilde s\rangle$ we also fix the behavior 
of  $\expval{P(w)}$ that is obtained from Eq.~\eqref{GuRMTgeral} using the inverse Fourier transform of Eq.~\eqref{chacfunc}.

\par
For high temperatures, in Fig. \ref{fig1} we see that the peak and the width of the work probability density, from single draws,
are close to $w^*=0$ and $\delta w=2N\expval{s}/\pi\approx 24.5$ according to our prediction in Eqs.(\ref{peakandwidth1}).
For low temperatures in Fig.\ref{fig3} we see that the peak and width of the work probability density, from single draws, are also
close to our predictions in Eqs.(\ref{peakandwidth2}): 
$w^*=24$ and $\delta w=2N\expval{\tilde s}/\pi=12.25$.
 

\section{Conclusions}
\label{SectionV}

Using random matrix theory we have developed a general description of the work characteristic function corresponding to 
the work probability density function for the so-called two-projective-energy-measurements scheme considering sudden quenches over
Gaussian ensembles of random Hamiltonians. When the Hamiltonian matrix dimension is large, our analytical expression for the ensemble average of 
the characteristic function describes the universal behavior of the exact work characteristic function calculated with the eigenenergies and 
eigenvectors of the initial and final Hamiltonians that are single draws of two independent Gaussian ensembles in all range of temperatures
of the initial thermal state. In this way we verify the ergodic property that ensures that the fluctuations of the characteristic 
function calculated from single draws from the ensembles, around our RMT average expression, go to zero in the limit of infinite 
matrix dimensions. In the case of relative small matrix dimensions our expression describes the average behavior of the 
work characteristic function calculated from single draws of the ensembles.

It is well known that RMT describes quite well some statistical properties of quantum systems with classically 
chaotic counterparts (quantum chaotic systems). In this sense, our work is one step forward in the study of the possibility 
of the existence of a general behavior of  the work probability density function for the so called two-projective-energy-measurements 
scheme in sudden quenches in real quantum chaotic systems.

\acknowledgments{E.G.A., L.C.C., N.G.A. and F.T. acknowledge financial support from the Brazilian funding agencies CNPq, 
CAPES (PROCAD2013 project), FAPERJ, FAPEG and the National Institute of Science and Technology - Quantum Information.
D.A.W. and A.J.R. have received funding from CONICET (Grant No PIP 11220150100493CO) ANPCyT(PICT- 2016-1056 and PICT 2014-3711), 
UBACyT (Grant No 20020130100406BA).

\appendix

\section{}
\label{AppA}

Here we show the derivations of the averages in Eqs.\eqref{eqmultiples} and \eqref{medEumenos} that are valid
in the regime of the large matrice dimension, $N$, which means that the density of levels of the spectra are described by
the semicircle law, given in Eq. \eqref{sc-law}.

The averages \eqref{finaltext}) and \eqref{medEumenos} can be fitted in the same notation.
We can write them as $\expval{e^{\pm iu {\cal E}_{j}}}$, where the minus sign corresponds to the case 
${\cal E}_j\rightarrow E_n$ and the plus sign to the case 
${\cal E}_j\rightarrow \tilde E_m$. Then we have: 
\begin{eqnarray}
  \expval{e^{\pm iu{\cal E}_{j}}}& = &   \int d{\boldsymbol{\mathcal{E}}}\,
 e^{\pm iu{\cal E}_{j}}\,P(\boldsymbol{\mathcal{E}})= \nonumber\\
&=& 
\frac{1}{N}\int_{-\infty}^{\infty}d{\cal E}_j\;e^{\pm iu{\cal E}_j}\, \rho({\cal E}_j),
\label{mediaexp}
\end{eqnarray}
where we have used, in the last step, the definition of the level density given in Eq.\eqref{dens} .
We now replace $\rho({\cal E}_j)\rightarrow \rho_{N\gg 1}(a,x)$  by the expression given by Eq.\eqref{sc-law} in the last equation,
with $ x = {\cal E}_j-\expval{\cal E}$ and $a = 2N\expval{s^\prime}/\pi$.
We have $\expval{s^\prime} = \expval{s}$,  $\expval{\cal E} = \expval{E}$ in the case ${\cal E}_j\rightarrow E_n$,
and $\expval{s^\prime} = \expval{\tilde{s}}$, $\expval{\cal E} = \langle\tilde{E} \rangle$ in the case ${\cal E}_j\rightarrow \tilde E_m$.
In doing that we reach  
\bea
\expval{e^{\pm iu{\mathcal{E}}_{j}}} & = & 
\frac{2}{\pi}\;e^{\pm iu\langle{\cal E}\rangle}\;\int_{-1}^{1}dy^\prime \;e^{\pm i u a\;y^\prime}\sqrt{1-(y^\prime)^{2}} \nonumber\\
&=&2\,e^{\pm iu\langle{\cal E}\rangle}\;\frac{J_1(a  u)}{a  u},
\label{eq:valf}
\eea
where we have defined the new variable $\;y^\prime = \frac{{\cal E}_j-\expval{\cal E}}{a}$,
and we have used that
\begin{equation}
\int_{-1}^{1}d{x}\;e^{\pm ic_{1}x}\;\sqrt{1 - x^{2}}= \pi\frac{J_{1}(c_{1})}{c_{1}}\;,
\end{equation}
where $J_{n}(x)$ is a Bessel function of the first kind (with $n=1$ in our case).

Analogous procedures provide the results of Eqs. \eqref{avpartition} and \eqref{avpartitionbeta}. In the former we have 
\bea
\expval{e^{-\beta E_{n}}} &=&
\frac{2}{\pi}\;e^{-\beta \langle E\rangle}\;\int_{-1}^{1}dy^\prime\;e^{- \beta a \;y^\prime}\sqrt{1-(y^\prime)^{2}} = \nonumber\\
&=& 2\,e^{-\beta \langle E \rangle}\;\frac{I_1(a  u)}{a  u},
\label{eq:z}
\eea
and in the latter, we have
\bea
\expval{e^{-\beta E_{n}}e^{-iuE_{n}}} &=&
\frac{2}{\pi}\;e^{-\beta \langle E\rangle}\;e^{-iu \langle E\rangle} \times \nonumber \\
&\times& \int_{-1}^{1}dy^\prime \;e^{-a(\beta + iu)\;y^\prime }\sqrt{1-(y^\prime )^{2}} = \nonumber\\
&=& e^{-\beta \langle E \rangle}\;e^{-iu \langle E\rangle}\;_{0}F_{1}\left[2\;, - \frac{a^2}{4}(u - i\beta)^2 \right], \nonumber \\
\label{eq:z}
\eea
where we defined $a = 2N\expval{s}/\pi$ and $\;y^\prime = \frac{ E_n -\expval{E}}{a}$
and used the results
\begin{equation}
\int_{-1}^{1}d{x}\;e^{-c_{2}x}\;\sqrt{1 - x^{2}}= \pi\frac{I_{1}(c_{2})}{c_{2}}
\end{equation}
and
\begin{equation}
\int_{-1}^{1}d{x}\;e^{-c_{3}(c_4 + ic_5)x}\;\sqrt{1 - x^{2}} = 
\frac{\pi}{2} \;_{0}F_{1}\left[2\;, -\frac{c_{2}^2}{4}(c_5 - ic_4)^2 \right], 
\end{equation}
with $ I_n(x)$ being a modified Bessel function of the first kind (with $n=1$ in our case) and 
$_{0}F_{1}\left[c,x\right]$ being the
confluent hypergeometric function.

\bibliographystyle{apsrev}

\end{document}